# Static Address Generation Easing: a Design Methodology for Parallel Interleaver Architectures

C. Chavet[2], P. Coussy[2], P. Urard[1] and E. Martin[2]

[1]STMicroelectronics

[2]Lab-STICC Lab., Université de Bretagne Sud

*Abstract*—**For high throughput applications, turbo-like iterative decoders are implemented with parallel architectures. However, to be efficient parallel architectures require to avoid collision accesses i.e. concurrent read/write accesses should not target the same memory block. This consideration applies to the two main classes of turbo-like codes which are Low Density Parity Check (LDPC) and Turbo-Codes. In this paper we propose a methodology which finds a collision-free mapping of the variables in the memory banks and which optimizes the resulting interleaving architecture. Finally, we show through a pedagogical example the interest of our approach compared to state-of-the-art techniques.**

## 1. INTRODUCTION

In the multimedia and telecommunications domain, continuously emerging customer services require severe performance to implement the new communication standards. Indeed, communication systems require high throughput -on the order of several hundred Mb/s- accompanied by both low latency and severe bit error rate BER constraints (e.g. wireless, fiber-optic communication…). Owing to their impressive near-Shannon-limit error correcting performance, turbo-like codes in their parallel or serially concatenated versions [3], originally dedicated to channel coding, or LDPC codes [4], are being currently reused in most of digital communication systems (e.g. equalization, demodulation, synchronization, MIMO…).

These coders are formed by two or more processing elements PE (encoders/decoders) and one communication network composed of steering components (multiplexers, butterflies, barrel shifters…) and memory elements (registers, RAMs…). This network interleaves the data blocks exchanged by the PEs according to a predefined rule named interleaving law or permutation law. The turbo decoding principle is based on an iterative algorithm using decoders exchanging information in order to improve the error correction performance through the iterations. The iterative nature of these algorithms is a severe constraint to satisfy the aforementioned requirements with an affordable implementation complexity. A widespread solution is to realize the turbo decoder in a parallel fashion. One the one hand, this solution increases the throughput since the latency of the system becomes the latency of constituent sub-blocks [3]. On the other hand, the complexity and the cost of the system are increased due to parallel nature of the architecture.

By the way, depending on the interleaving law, different parallel processing elements may try to simultaneously access the same memory block (cf. Fig. 1). This problem is known as the "collision" problem [7]. In this case, three classes of solution are available: The designer may:

- define his own dedicated interleaving law in order to avoid such collision problems, but the resulting architecture may not be standard compliant.
- add extra memory elements and control logic in the communication network in order to buffer and postpone the conflicting data.
- find a memory mapping avoiding any conflict access while taking into account the cost of the architecture (i.e. of the communication network).

The paper is organized as follows: the second section presents the existing solutions to design parallel interleaver architectures. The third section is dedicated to the problem formulation of the interleaver design. In the fourth section we present the approach we propose to automatically find a memory mapping solution that avoids any conflict access. Finally, the last section presents experimental results on a pedagogical example.

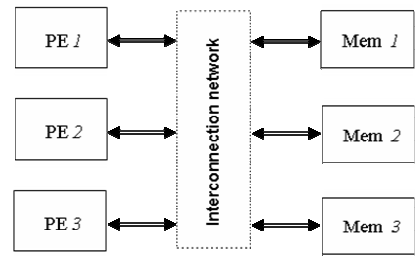

**Fig. 1:** *Memory collision problem*

## 2. RELATED WORKS

Interleaving law is a permutation law, also referred as $\Pi$, that scrambles data to break up neighbourhood-relations [7]. It is a key factor for turbo-codes performances, which varies from one communication standard to another. Moreover within a given standard, different interleaving rules can be used for different modes through varying frame lengths and/or data rates [5]. In this context, taking into account the aforementioned constraints and the collision problems to design hardware implementations of parallel turbo decoders require the integration of complex interconnection network topology (cf. Fig. 1) supporting the intensive interleaved memory accesses. Indeed, in state-of-the-art parallel turbo-decoding, interleaving is considered as a limiting factor concerning the overall system performance and the architectural cost.

To successfully tackle these problems, different solutions have been recently proposed.

A first solution to get rid of collisions with nonprunable interleavers, consists in designing a specific interleaver rule. In [7], the authors propose a deterministic methodology to design collision-free interleavers. In [8] and [6] the authors define collision-free permutations thanks to a combination of a spatial and a temporal permutation. The authors of [9] simply integrate the collision-free constraint in the design of their interleaver. However, the multi-modes architectures (depending on the frame length, the data-rate…) can not be handled by such approaches. Another solution consists in defining a collision-free interleaver that preserves this property even when pruned. In [5], the authors describe a design rule to obtain such interleavers, with an incremental algorithm that generates collision-free interleavers by adding new elements in successive steps, to a small initial permutation. Of course, all these solutions are viable if and only if the designer is free to choose the permutation law to be used in the system. As a consequence, the resulting architecture may not be standard compliant.

A second approach consists in adding extra memory elements in the communication network. The aim is to buffer and to postpone the conflicting data. In [1] the authors propose, when a collision appears, to store the conflicting information in the communication network until the targeted sub-block can process it. Of course, the additional network buffering resources, and consequently the time needed to interleave information, increase with the number of parallel processors. This is a suboptimal strategy, in terms of latency and thus throughput, which avoids collisions at the expense of area and memory. Moreover, the communication is based on a Benes network [2], which might be suboptimal compared to a dedicated and optimized architecture. Unlike these implementations, in [10] the authors propose a solution based on software and/or reconfigurable parts to achieve the required flexibility, but achieving lower throughput. In [11], an advanced heterogeneous communication network implementation was proposed. Two multistage interconnection network architectures are presented in order to handle on-chip communications in multiprocessor parallel turbo decoders. They are based on a dedicated network and associated routers. The main feature of these network architectures (Butterfly and Benes based topologies) is their supposed scalability enabling seamless trade-off between hardware complexity and available bandwidth for turbo decoding. The Butterfly network, which lacks of diversity, is a multistage interconnection network with 2-input 2-output routers. There is a unique path between each source and destination. As a consequence, the risk of conflict is increased and the authors have to add queues to store conflicting information. The second network architecture proposed is based on a Benes network. In this case, the latency is constant for all the couples (source, destination), but this network avoids the conflicts if and only if all the paths have a different destination. Unfortunately, it has been shown that it was not true for turbo-decoding applications because interleaving (respectively de-interleaving) ends in potential conflicts. Moreover, as already mentioned the Benes networks are costly and under-optimized solutions. In [12] the authors propose another on-chip interconnection network adapted to a flexible multiprocessor LDPC decoder based on the de Bruijn network. This network allows to efficiently supporting the communication intensive nature of the application. The conflict access are avoided thanks to a dedicated routing algorithm.

A third solution consists in finding a memory mapping avoiding any conflict access. Hence, the authors of [13] describe an approach that avoids collisions for every interleaver and any degree of parallelism. Contrary to the literature belief, the author have proven that for any code and any read/write operations scheduling, there exist a suitable memory mapping that grants a collision-free access. This solution automatically finds a collision-free data memory mapping respecting the interleaving rule, thanks to a simulated-annealing algorithm. As a consequence, the user cannot predict when the algorithm will end. Moreover, the proposed approach neither targets the optimization of the storage elements, nor the optimization of the network.

Finally some solutions based on a set of elementary memorising elements (Registers, FIFO, LIFO), such as [15], have been proposed. But if these solutions are able to generate strongly optimized architectures, they can not, to this day, target memory block based architecture.

In this paper, we present our patented approach named S.A.G.E.[1] (Static Address Generation Easing) dedicated to the memory mapping in block-based and parallel interleaver architectures. Counter to previous work, the proposed method considers both the generation of a conflict-free in-place memory mapping for any interleaving law (as well as [13] or [1]) and it is able to optimize the interconnection network (as well as [8]) in order to target a specific steering component to compose an optimized interconnection network between the PEs and the memory banks (if the interleaving rule enables to use this steering component, e.g. a barrel-shifter, a butterfly...).

## 3. PROBLEM FORMULATION

Let us consider a set of $L$ elements $E = \{e_1,... e_L\}$. Suppose we are given two different partitions on $E$, namely: **Nat = $\{E_1...E_N\}$** and **Int= $\Pi$ = $\{E'_1...E'_N\}$**. These partitions have the following characteristics: all subsets $E_i$, $E'_i$ ($i = 1,...,N$), have the same number of elements $|E_i| = X = L/N$. Note that $N$ must be a divisor of $L$. In other words, a set $E_i$ (resp. $E'_i$) represents the data processed at the same time $i$ for the partition *Nat* (resp. *Int*). $N$ is the number of cycles required to process all the data and $X$ is the resulting parallelism (number of memory banks and number of processing elements). The following definition defines the mandatory constraints to design a conflict-free architecture.

*Definition*: Let $E$, *Int*, and *Nat* be defined as above. A function $M:\{1,...,L\} \rightarrow \{1,...,X\}$ is a mapping function for (*Nat, Int*) if it satisfies the following conditions for every $i$, $i' = 1, ..., L$, $i \neq i'$.

$$e_i, e_{i'} \in E_i \text{ for some } i => M(i) \neq M(i') \quad (1)$$

$$e_i, e_{i'} \in E'_i \text{ for some } i => M(i) \neq M(i') \quad (2)$$

or in other words, elements belonging to the same subset in either partition are mapped to different memory block because they will be accessed at the same time. The mapping function gives the correspondence between the variables and the memory banks. If the constraints (1) and (2) are all satisfied, no collision in the memory access will take place.

An interleaver architecture is shown Fig. 1. In this pedagogical example, three processing elements compute data and store the results in three memory banks, through an interconnection network. The objective is to be able to compute a memory mapping which satisfies the constraints (1) and (2), and which also reduces the complexity of the interconnection network as much as the interleaving law allows it. A dedicated design approach is thus needed. This approach has to respect both the interleaving rule and the design constraints (parallelism, number of memory bank, size of the memory banks, latency, throughput…). In order to optimize the architecture, the approach has also to take into account the steering components the designer wants the interconnection network to be based on.

## 4. SAGE APPROACH

### A. Interleaving Law

As previously mentioned, an interleaver is a component that shuffles data. It means that from a given input data order *Nat*, referred as natural order in this paper (e.g., *Nat* = 0, 1, 2, 3, 4…), the architecture has to genetrate the data in a different output order, referred as the interleaved mode *Int* (e.g., *Int* = 1, 9, 10, 5, 0…). The problem is to be able to design the interleaving architecture.

In order to generate a valid memory mapping, the SAGE algorithm represents these two data ordering (both natural and interleaved orders) with two matrixes as shown in Fig. 2.

| 0 | 1 | 2 | 3 |
|---|---|---|---|
| 4 | 5 | 6 | 7 |
| 8 | 9 | 10 | 11 |

| 1 | 5 | 2 | 6 |
|---|---|---|---|
| 9 | 0 | 7 | 8 |
| 10 | 11 | 3 | 4 |

a- Natural order matrix - $M_{Nat}$  b- Interleaved order matrix – $M_{Int}$

**Fig. 2:** *SAGE reference matrixes*

In this example, the sets $E_i$ (resp. $E'_i$) are the columns of the matrix $M_{Nat}$ (resp. $M_{Int}$). The lines of the matrix refer to the processing

---

elements, e.g. the first line of each matrix refers to the data to be computed/stored by the same processing element: $PE_0$ in Fig. 1.

*B. Memory Mapping Constraints and Objectives*

There are two kinds of constraints/objectives to deal with: the structural constraints, which will guarantee the validity of the constraints (1) and (2); and the architectural objectives, which will be used to guide the memory mapping algorithm in order to implement the interconnection network based on specific steering components (e.g. a barrel-shifter based network). The structural constraints are mandatory in order to ensure the functional correctness of the resulting memory mapping. On the contrary, if the interleaving law intrinsically forbids to design the interleaver observing to the targeted architecture, then this objective may not be reached.

*C. SAGE Algorithm*

The SAGE algorithm uses two additional matrixes ($MAP_{Nat}$ and $MAP_{Int}$ in Fig. 3) in order represents the memory mapping. These two matrixes correspond respectively to $M_{Nat}$ and $M_{Int}$ and are initially empty.

| - | - | - | - |
|---|---|---|---|
| - | - | - | - |
| - | - | - | - |

a- Natural mapping - $MAP_{Nat}$

| - | - | - | - |
|---|---|---|---|
| - | - | - | - |
| - | - | - | - |

b- Interleaved mapping – $MAP_{Int}$

**Fig. 3:** *SAGE Mapping matrixes*

In matrix $MAP_{Nat}$ (resp. $MAP_{Int}$) each element (i, j) will be filled with a memory bank $b_i$. This will mean that the data in $M_{Nat}(i, j)$ (resp. $M_{Int}(i, j)$) will be stored in $b_i$. This memory mapping will be done according to aforementioned constraints.

Structural constraints:
- The memory mapping in $MAP_{Nat}$ and $MAP_{Int}$ for any data in common between $M_{Nat}$ and $M_{Int}$ must be the same.
- In any column of $MAP_{Nat}$ and $MAP_{Int}$ each memory has to be used only one time.

Architectural objectives:
- The memory mapping in a given column of $MAP_{Nat}$ (resp. $MAP_{Int}$) has to respect the rules of the steering components that compose the network.

For example, if we consider a barrel-shifter as a steering component, the memory mapping in a given column of $MAP_{Nat}$ (resp. $MAP_{Int}$) has to be a circular permutation of any other column of $MAP_{Nat}$ (resp. $MAP_{Int}$). The initialization of the SAGE mapping algorithm consists in assigning a memory bank for a first set of data, e.g. the first column of $MAP_{Nat}$ in Fig. 3.

Next, the corresponding data in the other matrix, $MAP_{Int}$ is updated with this mapping information. Once this update has been done, the SAGE algorithm selects the most constrained column (i.e. the most constraint cycle) and tries to find a memory mapping for the data which have not been assigned, with respect to structural constraints and architectural objectives. In order to do this, the algorithm constructs for all empty cells of the selected column a list of all available memory banks (respecting the structural constraints for the current column and the column of the other matrix in which the data of the current cell is stored). This list is ordered by taking into account the targeted architecture (the first elements are those which implement the targeted architecture). If a valid mapping is possible, i.e. all the lists generated for the current columns have at least one element, then the mapping is done with the first element of each list. Then, this mapping is reported in the second matrix and the recursion is performed. If one of the generated lists for the current column is empty, then this means that there is no solution with the current mapping. As a consequence the recursion is stopped and the algorithm goes back in order to select the next element of the list created at the previous iteration.

The resulting matrixes represents a conflict free memory mapping for the given interleaving law, and it also gives the control steps of the interleaving network.

```
En;         // Enable-boolean variable (Initialized with TRUE)
R ;         // Targeted architectural constraints
M_Nat;      // Natural mapping matrix
M_Int;      // Interleaved mapping matrix
T_MC_i;     // Targeted column C_i in matrix M (Natural or interleaved)
L_C_i;      // List of valid mapping solution for column C_i

if (En = TRUE) then
   // Search for the targeted column
   T_MC_i = Select_Target_Column (M_Nat, M_Int);
   // Generation of the list of memory mappnig
   L_C_i = Valid_Mapping_Solution(M_Nat, M_Int, T_MC_i);
   if (L_C_i is empty) then
       return (En = FALSE)
   else
       L_C_i = Mapping_Solution_Ordering (L_C_i, R);
       // Map the first solution in the list L_C_i = {l_1, l_2...}
       Affect_&_Report(M_Nat, M_Int, T_MC_i , l_1);
       En = TRUE;
       MemMap(M_Nat, M_Int, T_MC_i, R, En, L_C_i);
   end if;
else
   // Remove the first element of the list of mapping solution
   L_C_i = L_C_i – l_1;
   if (L_C_i is empty) then
       return (En = FALSE)
   else
       // Map the first solution in the list L_C_i = {l_2...}
       Affect_&_Report(M_Nat, M_Int, T_MC_i , l_2);
       En = TRUE;
       MemMap(M_Nat, M_Int, T_MC_i, R, En, L_C_i);
   end if ;
end if ;
```

**Fig. 4**: *Memory mapping algorithms – MemMap function*

Our recursive algorithm is able to find a valid memory mapping, and each time the interleaving law enables it, this mapping will respect the input architectural objective.

## 5. PRATICAL IMPLEMENTATION

Let's take as an example the interleaving represented in Fig. 2, $\prod=\{1, 9, 10, 5, 0, 11, 2, 7, 3, 6, 8, 4\}$, with 3 *PEs* and a targeted barrel shifter based steering component.

Let's suppose that we use the approach presented in [13]. This approach is in fact the closest one to our methodology. The algorithm proposed by the author starts with a tiled matrix that represents the interleaved parallel accesses on the natural order matrix (see Fig. 5, in this figure the tiles are $T_0$, $T_1$..). If two data are accessed in the same time in the interleaved order, then they get the same tile. Then, this algorithm first fills a mapping matrix, see Fig. 6.a with a greedy algorithm: if a memory bank $b$ ($b$ may be A, B or C) is usable without any conflict, then use $b$ ; if there is no simple solution then keep the cell empty.

| $T_1$ | $T_0$ | $T_2$ | $T_2$ |
|---|---|---|---|
| $T_3$ | $T_1$ | $T_3$ | $T_2$ |
| $T_3$ | $T_0$ | $T_0$ | $T_1$ |

**Fig. 5:** *Initialization of the tiled matrix*

Two constraints are mandatory in this algorithm: if two data are in the same column (natural order access), or if they get the same tile (interleaved order access), then their memory banks must be different (This is similar to our structural constraints). Once this first mapping matrix has been generated, a simulated-annealing is next used to compute the final mapping: this algorithm forces a conflicting memory bank in one of the empty cell.

| A | A | A | - |
|---|---|---|---|
| B | B | - | B |
| C | C | B | C |

a- Naïve mapping

| C | A | A | C |
|---|---|---|---|
| B | B | C | B |
| A | C | B | A |

b- Final mapping

**Fig. 6:** *Mapping matrix*

This will create a conflict access, which will be solved, maybe creating a new conflict with another data in the matrix… By the way, all the generated conflicts will be solved step-by-step, and the algorithm will be able to fill another empty cell (see [13] for more details).

Even if this algorithm always finds a mapping solution, it has no control on the resulting architecture since the steering components are not taken into account during the mapping and Fig. 6 shows that a barrel-shifter based architecture can not be generated.

On the contrary, our SAGE algorithm is able to take this objective into account. The first step of the SAGE mapping algorithm consists in assigning a memory bank for a first set of data, e.g. the first column of $MAP_{Nat}$ in Fig. 7: $MAP_{Nat}(0)=A$, $MAP_{Nat}(4)=B$ and $MAP_{Nat}(8)=C$.

| A | - | - | - |
|---|---|---|---|
| B | - | - | - |
| C | - | - | - |

a- Natural mapping - $MAP_{Nat}$

| - | - | - | - |
|---|---|---|---|
| - | - | - | - |
| - | - | - | - |

b- Interleaved mapping– $MAP_{Int}$

**Fig. 7:** *Initialization of the mapping matrix*

Then, the corresponding data in the other matrix, $MAP_{Int}$ in our example, are updated with this mapping information. This can be seen in Fig. 8 where the reported memory banks are in bold italic.

| A | - | - | - |
|---|---|---|---|
| B | - | - | - |
| C | - | - | - |

a- Natural mapping - $MAP_{Nat}$

| - | - | - | - |
|---|---|---|---|
| - | ***A*** | - | ***C*** |
| - | - | - | ***B*** |

b- Interleaved mapping– $MAP_{Int}$

**Fig. 8:** *Memory mapping transfer*

Once this update has been done, the SAGE algorithm selects the most constrained column (i.e. the most constraint cycle) and tries to assign a memory mapping with respect to both the structural and the architectural constraints. In Fig. 9, the most constrained column is the last column of $MAP_{Int}$ and there is only one mapping solution: $MAP_{Int}(6)=A$, and this mapping is reported in $MAP_{Nat}(6)=A$.

| A | - | - | - |
|---|---|---|---|
| B | - | ***A*** | - |
| C | - | - | - |

a- Natural mapping - $MAP_{Nat}$

| - | - | - | ***A*** |
|---|---|---|---|
| - | A | - | C |
| - | - | - | B |

b- Interleaved mapping– $MAP_{Int}$

**Fig. 9:** *Column selection*

Then our algorithm is performed on the rest of the matrixes: the most constrained column may be the third column of $MAP_{Nat}$ for example. In this case, in order to respect the objective of a barrel-shifter based architecture, the memory mapping must be $MAP_{Nat}(2)=C$ and $MAP_{Nat}(10)=B$. Indeed, C-A-B is the only circular shift of the reference column in this matrix (i.e. A-B-C) in this case.

| A | - | ***C*** | - |
|---|---|---|---|
| B | - | A | - |
| C | - | ***B*** | - |

a- Natural mapping - $MAP_{Nat}$

| - | - | ***C*** | A |
|---|---|---|---|
| - | A | - | C |
| ***B*** | - | - | B |

b- Interleaved mapping– $MAP_{Int}$

**Fig. 10:** *Memory mapping for barrel shifter*

Fig. 11.g and h, show a resulting valid memory mapping for the input constraint: *Bank A={0, 1, 6, 3}, Bank B={4, 5, 10, 7} and Bank C={8, 9, 2, 11}*. This solution enables the use of a barrel-shifter to implement the interconnection network since in each matrix, a column is always a circular permutation of any other column of the matrix. The mapping matrices also give the network control information: in natural access, there are only two control switch for memory access to/from the third column in $MAP_{Nat}$ (cf.Fig.11); in interleaved order, the barrel-shifter as to be switch from one column to another (The first column is A-C-B, then the second column, B-A-C, is a one step rotation of the first column…).

| A | ***A*** | C | - |
|---|---|---|---|
| B | - | A | - |
| C | ***C*** | B | - |

a- Natural mapping - $MAP_{Nat}$

| **A** | - | C | A |
|---|---|---|---|
| **C** | A | - | C |
| B | - | - | B |

b- Interleaved mapping– $MAP_{Int}$

| A | A | C | - |
|---|---|---|---|
| B | ***B*** | A | - |
| C | C | B | - |

c- Natural mapping - $MAP_{Nat}$

| A | ***B*** | C | A |
|---|---|---|---|
| C | A | - | C |
| B | - | - | B |

d- Interleaved mapping– $MAP_{Int}$

| A | A | C | - |
|---|---|---|---|
| B | B | A | - |
| C | C | B | ***C*** |

e- Natural mapping - $MAP_{Nat}$

| A | B | C | A |
|---|---|---|---|
| C | A | - | C |
| B | ***C*** | - | B |

f- Interleaved mapping– $MAP_{Int}$

| A | A | C | ***A*** |
|---|---|---|---|
| B | B | A | ***B*** |
| C | C | B | C |

g- Natural mapping - $MAP_{Nat}$

| A | B | C | A |
|---|---|---|---|
| C | A | ***B*** | C |
| B | C | ***A*** | B |

h- Interleaved mapping– $MAP_{Int}$

**Fig. 11:** *End of the SAGE algorithm*

## 5. CONCLUSION

In this paper, we have presented a memory mapping methodology named Static Address Generation Easing to design parallel interleaver architecture. This methodology allows to generate a valid memory mapping in any case, and if the interleaving law enables it, then the resulting memory mapping will respect the targeted interconnection network. Our approach has been compared through a pedagogical example to the state-of-the-art techniques and its interest has been shown